\documentclass[english,runningheads,11pt]{llncs}
\usepackage{tabularx,booktabs,multirow,delarray,array}
\usepackage{graphicx,amssymb,amsmath,amssymb}
\usepackage{latexsym}

\usepackage[in]{fullpage}

\usepackage{lineno}

\def\calP{\mathcal{P}}
\def\calD{\mathcal{D}}
\def\calT{\mathcal{T}}
\def\st{$s$-$t$}

\newcommand{\hd}{\mbox{$H\!D$}}
\newcommand{\vd}{\mbox{$V\!D$}}

\newcommand{\anc}{a}

\begin{document}
%\linenumbers

\title{$L_1$ Shortest Path Queries in Simple Polygons}

\author{Sang Won Bae\inst{1}
\and
Haitao Wang\inst{2}
%\thanks{H.~Wang's research was supported in part by NSF under Grant CCF-1317143.}
}

 \institute{
 Department of Computer Science,
 Kyonggi University, Suwon, South Korea\\
 \email{swbae@kgu.ac.kr} \\
\and
  Department of Computer Science,
  Utah State University, Logan, Utah, USA\\
  \email{haitao.wang@usu.edu}
}

\maketitle

\pagenumbering{arabic}
\setcounter{page}{1}

%\vspace*{-0.2in}
\begin{abstract}
Let $\calP$ be a simple polygon of $n$ vertices. We consider two-point
$L_1$ shortest path queries in $\calP$. We build a data
structure of $O(n)$ size in $O(n)$ time such that given any two query
points $s$ and $t$, the length of an $L_1$ shortest path from $s$ to $t$ in $\calP$
can be computed in $O(\log n)$ time, or in $O(1)$ time if both $s$
and $t$ are vertices of $\calP$, and an actual shortest path
can be output in additional linear time in the number of edges of the path.
To achieve the result, we propose a mountain decomposition of simple polygons, which may be interesting in its own right. Most importantly, our approach is much simpler than the previous work on this problem.
\end{abstract}


\section{Introduction}
\label{sec:intro}

Let $\calP$ be a simple polygon of $n$ vertices. We consider two-point shortest
path queries in $\calP$ for which the path lengths are measured in the $L_1$ metric.
The problem is to build a data structure to quickly compute an $L_1$ shortest path
or only compute the path length between any two query points $s$ and $t$.

If the Euclidean metric is instead used to measure the path lengths,
Guibas and Hershberger~\cite{ref:GuibasOp89} built a data structure of $O(n)$ space in $O(n)$ time that can answer each two-point Euclidean shortest path query in $O(\log n)$ time. It is well-known that a Euclidean shortest path in $\calP$ is also an $L_1$ shortest path~\cite{ref:HershbergerCo94}. Therefore, using the data structure in \cite{ref:GuibasOp89}, one can answer each two-point $L_1$ shortest path query in $O(\log n)$ time. However, since the data structure \cite{ref:GuibasOp89} is particularly for the Euclidean metric, both the data structure and the query algorithm are quite complicated.
Indeed, problems in the $L_1$ metric are usually easier than their
counterparts in the Euclidean metric. As a fundamental problem in
computational geometry, it is desirable to have a simpler approach for the two-point $L_1$ shortest path query problem.

In this paper, we present such a data structure of $O(n)$ space that
can be built in $O(n)$ time. With the data structure, given any two
query points $s$ and $t$, we can compute the length of an $L_1$ shortest path
from $s$ to $t$ in $\calP$ in $O(\log n)$ time.
Further, if $s$ and $t$ are both vertices of $\calP$, then the query
time of our data structure becomes only $O(1)$.
An actual shortest path can be output
in additional time linear in the number of edges in the path.

Our method has several advantages compared to \cite{ref:GuibasOp89}.
First, our method is much simpler.
Second, if both $s$ and $t$ are
vertices of $\calP$, then the query time of our algorithm is only $O(1)$,
while the data structure in~\cite{ref:GuibasOp89} still needs $O(\log n) $ query time.
In addition, using our techniques, we can obtain the
following result. Given a set $S$ of $m$ points in $\calP$, we can
build a data structure of $O(n+m)$ size in $O(n+m\log n)$ time such
that each two-point $L_1$ shortest path query can be answered in $O(1)$ time for any two query points in $S$.

Given a point $s$ in $\calP$ as a single source, a Euclidean shortest
path map in $\calP$ can be built in $O(n)$ time \cite{ref:GuibasLi87}
and the map is usually used for answering single-source shortest path
queries (i.e., $s$ is fixed and only $t$ is a query point). Again, as a Euclidean shortest path in $\calP$ is also an $L_1$ shortest
path, the Euclidean shortest path map of $s$ is also an
$L_1$ shortest path map. As a by-product of our techniques, we present
another simpler way to compute an $L_1$ shortest path map for $s$ in $O(n)$
time.
%and our approach is also quite simple.
%r than the Euclidean one in~\cite{ref:GuibasLi87}.

Many previous results, e.g., \cite{ref:BaeL116,ref:BaeCo15,ref:ChenTw16},
have normally resorted to the corresponding Euclidean data structures
\cite{ref:GuibasOp89,ref:GuibasLi87}
in order to answer $L_1$ shortest path queries in simple polygons.
With our simpler (and faster, if both query points are vertices of $\calP$ in the two-point queries) solutions, the algorithms in those previous work can be simplified as well.

\subsection{Our Techniques}

If $\calP$ is a simple {\em rectilinear} polygon where every edge of $\calP$ is parallel to either the $x$- or $y$-axis, then Schuierer~\cite{ref:SchuiererAn96} gave an $O(n)$-size data structure that can be built in $O(n)$ time such that given any two query points $s$ and $t$ in $\calP$, a shortest rectilinear shortest \st\ path can be found in $O(\log n)$ time, and if both $s$ and $t$ are vertices of $\calP$, then the query time becomes $O(1)$.

Our approach follows a similar scheme as the one in \cite{ref:SchuiererAn96} but extends the result in \cite{ref:SchuiererAn96} for simple rectilinear polygons to arbitrary simple polygons. Specifically, a main geometric structure used in \cite{ref:SchuiererAn96} is a histogram partition, which is commonly used for solving problems in simple rectilinear polygons, e.g., see~\cite{ref:SchuiererCo94}. We generalize the concept to arbitrary simple polygons and develop a so-called {\em mountain decomposition} for arbitrary simple polygons. The mountain decomposition may be interesting in its own right and find other applications on simple polygons as well.

The rest of the paper is organized as follows. In
Section~\ref{sec:decom}, we introduce the mountain decomposition of
$\calP$. In Section~\ref{sec:spm}, we show how an $L_1$ shortest path
map can easily be obtained by using the mountain decomposition. In
Section~\ref{sec:twopoint}, we solve the two-point $L_1$ shortest path
query problem. Section~\ref{sec:applications} presents some
applications of our new results.

In the following, unless otherwise stated, a shortest path
refers to an $L_1$ shortest path in $\calP$ and the path
length is measured in the $L_1$ metric. For ease of exposition,
we make a general position assumption that no two vertices of $\calP$
have the same $x$- or $y$-coordinate (otherwise we could slightly perturb the data to achieve the assumption).
%The assumption can be lifted without affecting the performance of our results, although the discussion would be more tedious.

\section{The Mountain Decomposition}
\label{sec:decom}

In this section, we introduce a decomposition of $\calP$, which we call the {\em mountain decomposition}. It generalizes the histogram partition of simple rectilinear polygons~\cite{ref:SchuiererAn96}. We first introduce some notation and concepts.

\subsection{Preliminaries}

For any two points $p$ and $q$ in $\calP$, let $\overline{pq}$ denote the line segment joining $p$ and $q$.
We use $d(p,q)$ to denote the length of an $L_1$ shortest path from $p$ to $q$ in $\calP$.
For convenience, sometimes we use $|pq|$ to refer to $d(p,q)$, if $d(p,q)$ is equal to the $L_1$ length of the line segment $\overline{pq}$ (which may not be entirely in $\calP$).
%If $\overline{pq}$ is contained in $\calP$, we use $|\overline{pq}|$ to denote its $L_1$ length.

%To prove some properties of shortest paths in the paper, we often use the following observation.
%Let $p_1$ and $p_2$ be two points in $\calP$. Let $q$ be the intersection of the vertical line through $p_1$ and the horizontal line through $p_2$. The observation is that if $\overline{p_1q}$ and $\overline{p_2q}$ are both in $\calP$, then $\overline{p_1q}\cup \overline{p_2q}$ is a shortest path from $p_1$ to $p_2$.

The {\em horizontal trapezoidal decomposition} $\hd(P)$ of a simple polygon $P$ is
a decomposition of $P$ into cells by extending a horizontal line
from each vertex of $P$ to the polygon interior until both of its ends hit the boundary of $P$~\cite{ref:ChazelleTr91}.
The {\em vertical trapezoidal decomposition} $\vd(P)$ is defined analogously.
Both decompositions can be computed in linear time~\cite{ref:ChazelleTr91}.

\begin{figure}[t]
\begin{minipage}[t]{0.49\linewidth}
\begin{center}
\includegraphics[totalheight=1.0in]{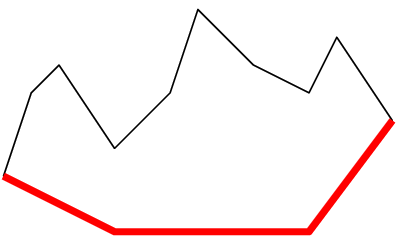}
\caption{\footnotesize Illustrating an upward mountain. The bottom is shown with thick (red) segments, and the horizontal edge is the base.
}
\label{fig:mountain}
\end{center}
\end{minipage}
\hspace*{0.05in}
\begin{minipage}[t]{0.49\linewidth}
\begin{center}
\includegraphics[totalheight=1.0in]{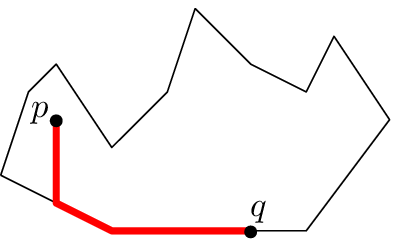}
\caption{\footnotesize Illustrating the canonical path $\pi_P(p,q)$ in an upward mountain.
}
\label{fig:canonical}
\end{center}
\end{minipage}
\vspace*{-0.15in}
\end{figure}

Next, we introduce a new concept, namely, {\em mountains}.
A polygonal chain $\gamma$ is {\em $x$-monotone} if the intersection of $\gamma$ with every vertical line is connected; $\gamma$ is called {\em $y$-monotone} if the intersection of $\gamma$ with every horizontal line is connected. A simple polygon $P$ is called an {\em upward mountain} if the following are satisfied (e.g., see Fig.~\ref{fig:mountain}): (1) the leftmost and rightmost vertices of $P$ divide the boundary of $P$ into two $x$-monotone chains; (2) the lower chain has a horizontal edge, called the {\em base}; (3) the lower chain has at most one edge on the left of the base, which is of negative slope and is called the {\em left-wing}; (4) the lower chain has at most one edge on the right of the base, which is of positive slope and is called the {\em right-wing}. According to the definition, the lower chain of $P$ has at most three edges, and we call the lower chain the {\em bottom} of $P$. If we rotate $P$ by $180^{\circ}$, then $P$ becomes a {\em downward  mountain}.
%(but its left-wing, right-wing, and bottom refer to the same things as before).
Similarly, we can define the {\em rightward} and {\em leftward mountains}, i.e., by rotating $P$ by $90^{\circ}$ and $270^{\circ}$ clockwise, respectively.

%For an $x$-monotone simple polygon $P$, if $P$ has a horizontal edge $e$ such that every vertical line that intersects $P$ also intersects $b$, then we call $P$ a {\em horizontal mountain} and call $e$ the {\em base} of $P$. See Fig.~\ref{fig:mountain} for an example.
%If a horizontal mountain $P$ is above (resp. below) its base, then $P$ is
%called an {\em upward} (resp., {\em downward}) mountain.
%The {\em vertical mountain} and {\em leftward/rightward mountain} are defined similarly.
%Note that a mountain may also be a triangle. However, in the following, since triangles are handled separately from the more general mountains, for convenience, a mountain will always refer to one that is not a triangle, unless stated otherwise.

Consider an upward mountain $P$. Let $p$ and $q$ be two points in $P$ with $q$ at the base of $P$. We call the following path from $p$ to $q$ the {\em canonical path} (e.g., see Fig.~\ref{fig:canonical}): From $p$, move vertically down until the bottom of $P$, then move to reach $q$ along the bottom. Let $\pi_P(p,q)$ denote the canonical path. Observe that $\pi_P(p,q)$, which has at most three edges, is both $x$- and $y$-monotone, and thus is a shortest path with length equal to $|pq|$.

\begin{figure}[h]
\begin{minipage}[t]{\linewidth}
\begin{center}
\includegraphics[totalheight=1.0in]{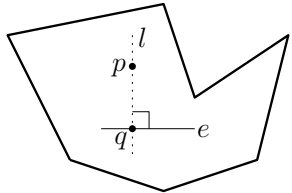}
\caption{\footnotesize The point $p$ is perpendicularly visible to $e$ and $q$ is the orthogonal projection of $p$ onto $e$.
}
\label{fig:projection}
\end{center}
\end{minipage}
\vspace*{-0.15in}
\end{figure}

For a horizontal or vertical line segment $e$ in $P$,
we say that a point $p\in P$ is {\em perpendicularly visible} to $e$
if the line $l$ through $p$ and orthogonal to $e$ intersects $e$ and the line segment connecting the intersection and $p$ is in $P$ (e.g., see Fig.~\ref{fig:projection}), and we call the intersection $l\cap e$ the {\em orthogonal projection} of $p$ onto $e$.

\subsection{The Mountain Decomposition}

Let $\calP$ be the given simple polygon of $n$ vertices, and
let $\partial\calP$ denote the boundary of $\calP$.
In the following, we introduce a {\em mountain decomposition} of $\calP$.

Our decomposition starts with an arbitrary horizontal line segment $e=\overline{ab}$
contained in $\calP$ with $a, b \in \partial \calP$.
%For convenience of discussion, the endpoints $a$ and $b$ of $e$ are considered to be vertices of $\calP$.
The segment $e$ partitions $\calP$ into two sub-polygons:
one locally above $e$ and the other locally below $e$; let $\calP_1$ and $\calP_2$ denote the former and the latter sub-polygons, respectively.
We only describe how to decompose $\calP_1$,
and the decomposition of $\calP_2$ is done analogously.
Roughly speaking, the decomposition partitions $\calP_1$ into mountains. The details are given below.

If $\calP_1$ is a triangle, which is a special mountain, then we are done with the decomposition and we call $e$ the {\em base} of the triangle. We use $M_e$ to denote this triangle.

\begin{figure}[t]
\begin{minipage}[t]{0.49\linewidth}
\begin{center}
\includegraphics[totalheight=1.6in]{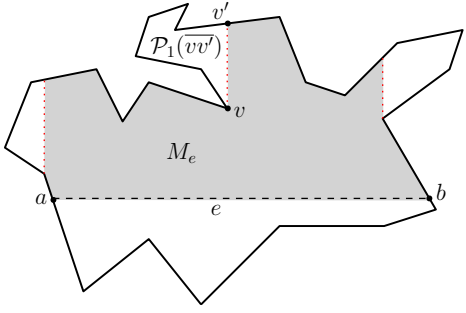}
\caption{\footnotesize Illustrating the mountain $M_e$, the gray region. $M_e$ has three windows, shown with dotted vertical segments.
}
\label{fig:mountaindecom}
\end{center}
\end{minipage}
\hspace*{0.02in}
\begin{minipage}[t]{0.49\linewidth}
\begin{center}
\includegraphics[totalheight=1.6in]{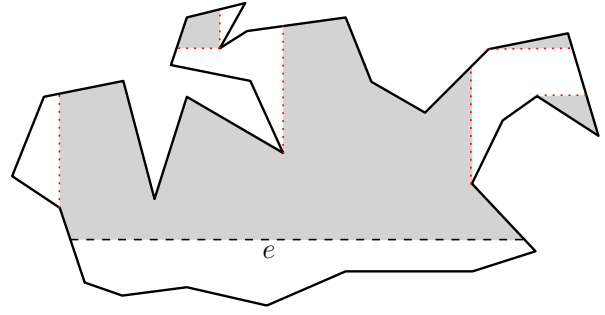}
\caption{\footnotesize Illustrating the mountain decomposition $\calD_1$ of $\calP_1$, i.e., the sub-polygon of $\calP$ above $e$.
}
\label{fig:fulldecom}
\end{center}
\end{minipage}
\vspace*{-0.15in}
\end{figure}

If $\calP_1$ is a not a triangle, let $M_e$ be the maximal mountain whose base is  $e$ (e.g., see Fig.~\ref{fig:mountaindecom}). Specifically, $M_e$ is defined as follows.
Without loss of generality, assume $a$ and $b$ are left and right endpoints of $e$, respectively.
If $a$ is in the relative interior of an edge of $\calP$, then define $a'$ to be the upper endpoint of the edge; otherwise, if $a$ is a vertex of $\calP$, then only one of the two adjacent vertices of $a$ belongs to $\calP_1$ and we let $a'$ refer to that vertex. In either case, $a'$ is the next clockwise vertex of $\calP_1$ from $a$.
The left-wing of $M_e$ is $\overline{aa'}$ if $\overline{aa'}$ is of negative slope; otherwise, $M_e$ has no left-wing.
%The right-wing of $M_e$ is defined similarly (but with respect to the positive slope).
Similarly, let $b'$ be the next counterclockwise vertex of $\calP_1$ from $b$. The right-wing of $M_e$ is $\overline{bb'}$ if $\overline{bb'}$ is of positive slope; otherwise, $M_e$ has no right-wing.
The wings (if any) and the base $e$ of $M_e$ together constitute the bottom of $M_e$. Let $v$ be a vertex of $\calP_1$ not on the bottom of $M_e$ such that $v$ is vertically visible to the bottom of $M_e$ (i.e., there is a point $p$ on the bottom such that $\overline{vp}$ is vertical and in $\calP$) and the two edges of $\calP_1$ incident to $v$ are on the same side of the vertical line through $v$.
Let $v'$ be the first point on $\partial \calP_1$ hit by the vertically upward ray from $v$.
We call the segment $\overline{vv'}$ a {\em window}.
Note that $\overline{vv'}$ lies in $\calP_1$ and divides $\calP_1$ into two sub-polygons: one contains $e$ and the other does not. Let $\calP_1(\overline{vv'})$ denote the latter sub-polygon. See Fig.~\ref{fig:mountaindecom}. In addition, for each endpoint $v$ of the bottom of $M_e$,
if the vertically upward ray from $v$ lies locally inside $\calP_1$, then $\overline{vv'}$ is also a window, where $v'$ is the first point on $\partial \calP_1$ hit by the ray, and we define $\calP_1(\overline{vv'})$ similarly. The mountain $M_e$ is the sub-polygon of $\calP_1$ excluding $\calP_1(w)$
for all such windows $w$. The windows defined above are considered to be windows of $M_e$.
If there are no windows, then $M_e$ is equal to $\calP_1$ and we are done with the decomposition;
otherwise, for each window $w$, we decompose the sub-polygon $\calP_1(w)$ recursively with respect to $w$ (so the window $w$ will become the base of a mountain in $\calP_1(w)$).

We use $\calD_1$ to denote the resulting decomposition of $\calP_1$
after the preceding decomposition process finishes (e.g., see Fig.~\ref{fig:fulldecom}).

\paragraph{Remark.} Observe that the mountain $M_e$ contains at least one vertex of $\calP_1$ other than the two endpoints of $e$. Indeed, if $M_e$ has at least one wing, then this is obviously true because the upper endpoint of the wing is such a vertex. Otherwise, one can verify that at least one vertex of $\calP_1\setminus\{e\}$ is vertically visible to $e$ and thus is in $M_e$. This property is crucial for proving the combinatorial size of $\calD_1$ later in Lemma~\ref{lem:D1}. This is also the reason we introduce the wings for mountains.
\vspace{0.1in}

The mountain decomposition of $\calP_1$ induces a natural tree structure, called the {\em mountain decomposition tree}, denoted by $\calT_1$. Each cell of $\calD_1$ corresponds to a node in $\calT_1$.
Consider any cell $C$ of $\calD_1$, which is a mountain. $C$ has a base $b$, which is also a window of a mountain unless $b=e$. If $b=e$, then we call $C$ the {\em root cell} of $\calD_1$.
Specifically, the root cell $C$ corresponds to the root of $\calT_1$.
For each window $w$ of $C$, $C$ has a child subtree corresponding to the mountain decomposition of $\calP_1(w)$.
Note that each leaf of $\calT_1$ corresponds to a mountain with no windows.

In the following, we sometimes use the cells of $\calD_1$ and the nodes of $\calT_1$ interchangeably.
Following the tree terminology, for each cell $C$ of $\calD_1$, we use {\em parent cell} to refer to the cell of $\calD_1$ corresponding to the parent node of $C$ in $\calT_1$.
The {\em ancestor} or {\em descendant cells} are defined similarly.
%; for convenience, we consider $C$ as an ancestor of itself.

\begin{lemma}\label{lem:D1}
The combinatorial size of $\calD_1$ is $O(n)$ and $\calD_1$ can be computed in $O(n)$ time.
\end{lemma}
\begin{proof}
To argue that the size of $\calD_1$ is $O(n)$, for each cell $C$ of $\calD_1$, as remarked above, other than the endpoints of the base of $C$, $C$ contains at least one vertex of $\calP_1$.
Observe that each vertex of $\calP_1$ is on the boundary of at most two cells of
$\calD_1$. Since $\calP_1$ has $O(n)$ vertices, the size of $\calD_1$ is $O(n)$.

%we further decompose each cell $C$ of $\calD_1$ as follows. If $C$ is a triangle, we do
%nothing. Otherwise, without loss of generality, we
%assume $C$ is an upward mountain. We decompose $C$ by
%its vertical trapezoidal decomposition.
%
%Let $\calD'_1$ refer to the new decomposition of $\calP_1$ as above.
%Clearly, $\calD'_1$ is a refinement of $\calD_1$. Hence, the size of $\calD'_1$ is at least
%that of $\calD_1$. Each cell of $\calD_1'$ is either a triangle or a
%trapezoid, and thus is of constant size. As remarked above, other than the endpoints of the base of $C$, $C$ contains at least one vertex of $\calP_1$. This also implies that the boundary of
%each cell of $\calD_1'$ contains at least one vertex of $\calP_1$.
%Further,
%each vertex of $\calP_1$ is on the boundary of at most three cells of
%$\calD_1'$. Since $\calP_1$ has $O(n)$ vertices,
%the size of $\calD_1'$ is $O(n)$ and the size
%of $\calD_1$ is also $O(n)$.

In the following, we present an $O(n)$ time algorithm to compute $\calD_1$.
We make use of the vertical trapezoidal decomposition $\vd(\calP_1)$ and the horizontal
trapezoidal decomposition $\hd(\calP_1)$. Both decompositions can be computed in
$O(n)$ time~\cite{ref:ChazelleTr91}.

In the beginning, we need to compute the mountain
$M_e$ (if it is not a triangle) and its windows. Let $B_e$ denote the bottom of $M_e$, which can be determined in constant time given the segment $e$.
Observe that the mountain $M_e$ is exactly the union of the cells of $\vd(\calP_1)$ that {\em
properly} intersect the bottom $B_e$ (i.e., the intersection contains
more than one point). Thus, starting from one endpoint of $B_e$, $M_e$
can be computed by traversing $\vd(\calP_1)$ along $B_e$ in $O(|M_e|)$
time, where $|M_e|$ is the combinatorial size of $M_e$. The above
traversing procedure can also identify the windows of $M_e$ simultaneously.

Let $w$ be any window of $M_e$. Note that an endpoint of $w$ must be a
vertex of $\calP_1$ and let $v$ denote that vertex (e.g., see Fig.~\ref{fig:mountaindecom} in which $w=\overline{vv'}$).
Without loss of generality, we assume the sub-polygon $\calP_1(w)$ is locally on the
left side of $w$. Let $M_w$ denote the mountain in $\calP_1(w)$ with
base $w$, and let $B_w$ denote the bottom of $M_w$, which can be determined in constant time since $w$ is known. Observe that $M_w$ is the union of the cells of
$\hd(\calP_1)$ that properly intersect $B_w$, restricted to lie on the left side of
$w$. Hence, starting from an endpoint of $B_w$, $M_w$ can be computed by
traversing $\hd(\calP_1)$ along $B_w$. In this way,
the mountain decomposition of the sub-polygon $\calP_1(w)$ can be
computed recursively.

For the time analysis, notice that the time for computing $M_w$ is linear in the number of cells of $\hd(\calP_1)$ intersecting $w$ and the wings of $B_w$ (if any). Because the wings are on the boundary of $\calP_1$, the total number of cells of $\hd(\calP_1)$ (or $\vd(\calP_1)$) intersecting the wings of $B_w$ for all such windows $w$ in the entire algorithm is $O(n)$. On the other hand,
each cell of $\hd(\calP_1)$ intersecting $w$ can be visited at most twice in the entire algorithm (i.e., for computing $M_w$ of at most two windows $w$, which are in the same mountain facing each other).
Hence, the total time for computing $\calD_1$ is
%can be computed in linear time in the combinatorial size of $D_1$, which is
$O(n)$.
%The lemma thus follows.
\qed
\end{proof}

The mountain decomposition $\calD_2$ on $\calP_2$ is built in the same way,
which induces a mountain decomposition tree $\calT_2$.
Consequently, $\calD_2$ is also of $O(n)$ size and can also be computed in $O(n)$ time.
Let $\calD$ denote the decomposition of the whole $\calP$ induced by $\calD_1$ and $\calD_2$. Note that
$\calD$ can be considered to be the mountain decomposition of $\calP$
with respect to the chosen line segment $e$, and $\calD$ is uniquely determined once $e$ is fixed.

\section{The $L_1$ Shortest Path Map and Single-Source Shortest Path Queries}
\label{sec:spm}

In this section, based on the mountain decomposition, we present a simple way
to construct a shortest path map for a fixed source point $s$ in $\calP$.
A shortest path map for $s$ is a decomposition of $\calP$ that encodes
all shortest paths from $s$ to all other points in $\calP$.
%so that single-source shortest path queries can be processed efficiently.

Let $e$ be the maximal horizontal line segment through $s$
that lies in $\calP$.
We compute the mountain decomposition $\calD$ with respect to $e$.
Define $\calD_1$, $\calD_2$, $\calT_1$, $\calT_2$ in the same way
as before.

\begin{figure}[t]
\begin{minipage}[t]{0.49\linewidth}
\begin{center}
\includegraphics[totalheight=2.0in]{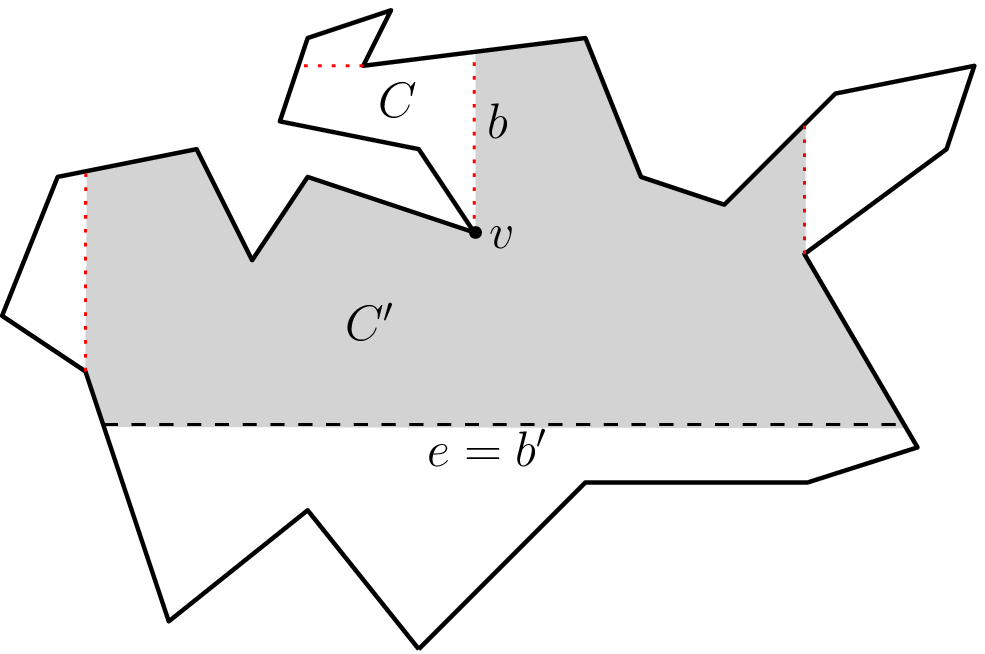}
\caption{\footnotesize Illustrating the definition of $\anc_C$: $e$ is $b'$ and $\anc_C$ is $v$.
}
\label{fig:root}
\end{center}
\end{minipage}
\hspace{0.02in}
\begin{minipage}[t]{0.49\linewidth}
\begin{center}
\includegraphics[totalheight=2.0in]{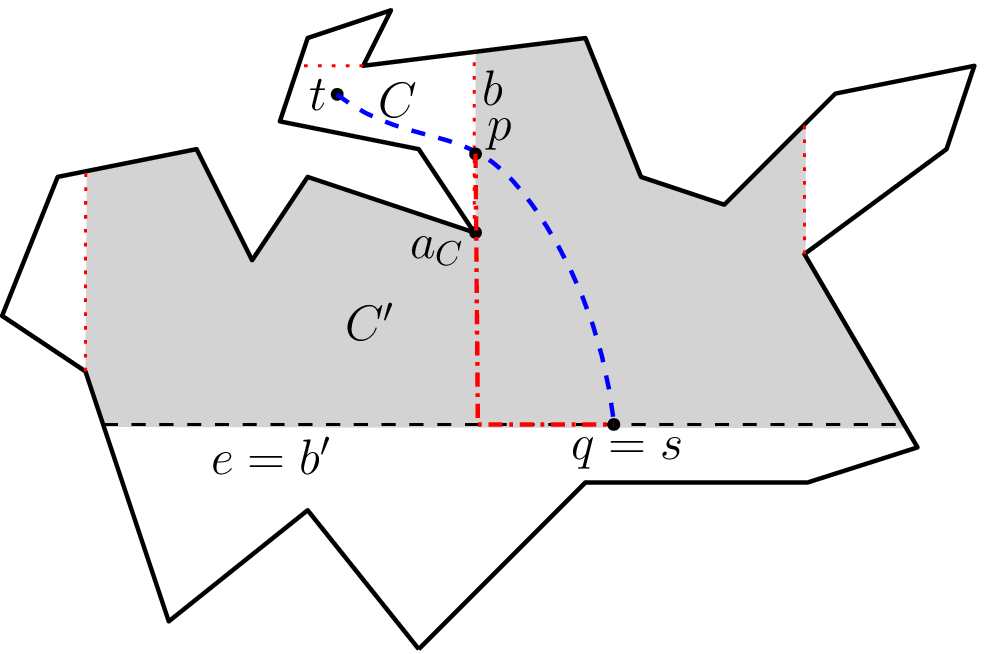}
\caption{\footnotesize Illustrating the proof of Lemma~\ref{lem:root}. The dashed (blue) path is a shortest path from $t$ to $s$. The dashed dotted (red) path is the canonical path $\pi_{C'}(p,q)$ from $p$ to $q$ in $C'$, which must contain the anchor $a_C$ of $C$.
}
\label{fig:path}
\end{center}
\end{minipage}
\vspace*{-0.15in}
\end{figure}

Consider any cell $C$ of $\calD$. Without loss of generality, we assume $C$ is in $\calD_1$.
We next define the {\em anchor} $\anc_C$ for $C$, which is a point on the base of $C$ and will be used for computing shortest paths from $s$ to all points in $C$.
If $C$ is the root cell (i.e., the base of $C$ is $e$), then $s$ is on the base of $C$ and we define $\anc_C$ to be $s$. Otherwise, let $b$ be the base of $C$.
According to our mountain decomposition, $b$ is a window of the parent cell $C'$ of $C$ in $\calD_1$ (e.g., see Fig.~\ref{fig:root}).
%Let $b'$ be the base of $C'$ (e.g., see Fig.~\ref{fig:root}).
Let $v$ be the endpoint of $b$ closer to the bottom of $C'$.
By our way of defining windows, $b$ is orthogonal to the base of $C'$ and $v$ must be a vertex of $\calP$. We define the anchor $\anc_C$ to be $v$.
Recall the definition of canonical paths in a mountain. Because $a_C\in b$, for each point $t\in C$, $C$ has a canonical path $\pi_C(t,a_C)$ from $t$ to $a_C$, which is a shortest path of length $|ta_C|$.
The following lemma explains why we need the anchors.

%We have the following lemma.

\begin{lemma}\label{lem:root}
Let $C$ be a cell of $\calD$.
For any point $t$ in $C$, there is a shortest \st\ path passing through the anchor $\anc_C$ and containing the canonical path $\pi_C(t,\anc_C)$.
%Further, if $C$ is a triangle, then there is a shortest \st\ path containing the segment $\overline{t\anc_C}$;
%if $C$ is a mountain, then there is a shortest \st\ path  containing $\overline{tt'}\cup \overline{t'\anc_C}$,
%where $t'$ is the orthogonal projection of $t$ onto the base $b$ of $C$.
\end{lemma}
\begin{proof}
If $C$ is the root cell, then since $s$ is on the base $e$ of $C$ and $\anc_C=s$, the lemma statement obviously holds.
%If $C$ is a triangle, then $\overline{st}$ is in $\calP$ and thus it is obviously true that $\overline{st}$ (which is $\overline{t\anc_C}$) is a shortest path.
%If $C$ is a mountain, then $\overline{tt'}$ is orthogonal to $\overline{t's}$, and thus
%$\overline{tt'}\cup \overline{t's}$ is a shortest path.
%Hence, the lemma holds if $C$ is the root cell.

In the following, we assume that $C$ is not the root cell.
This means that $C$ has a parent cell $C'$.
Let $b$ be the base of $C$ and $b'$ be the base of $C'$.
%According to the definition, $\anc_C$ is the endpoint of $b$ closer to $b'$.
Let $\pi$ be any shortest path from $s$ to $t$.
Since the base $b$ separates $C$ from $e$ and thus separates $t$ from $s$,
$\pi$ must intersect $b$ at a point $p$. See Fig.~\ref{fig:path}.
If $b'= e$, since $s\in e$, $\pi$ must intersect $b'$. Otherwise, $b'$ separates $C'$ from $e$ and thus $\pi$ also intersects $b'$.
In either case, let $q$ be an intersection between $\pi$ and $b'$.
%Let $q$ be the orthogonal projection of $p$ onto $b'$.
%Since $\overline{pq}$ is orthogonal to $\overline{qp'}$, $\overline{pq}\cup\overline{qp'}$ is a shortest path from $p$ to $p'$.
Since the canonical path $\pi_{C'}(p,q)$ is a shortest path from $p$ to $q$, by replacing the portion of $\pi$ between $p$ and $q$ by $\pi_{C'}(p,q)$, we can obtain another shortest \st\ path $\pi'$ that contains $\pi_{C'}(p,q)$. Notice that according to the definition of $\anc_C$ and the definition of $\pi_{C'}(p,q)$, $\anc_C$ must be on $\pi_{C'}(p,q)$; see Fig.~\ref{fig:path}. Thus, the shortest path $\pi'$ passes through $\anc_C$.

Further, since $\pi_C(t,\anc_C)$ is a shortest path from $t$ to $\anc_C$,
%if $C$ is a triangle, then $\overline{\anc_C t}$ is in $\calP$ and thus is a shortest path from $\anc_C$ to $t$, while
%if $C$ is a mountain, then $\overline{tt'}\cup \overline{t'\anc_C}$ is a shortest path from $\anc_C$ to $t$,
%since $\overline{tt'}$ is perpendicular to $\overline{t'\anc_C}$ and $t$ is visible to $b$.
if we replace the portion of $\pi'$ between $\anc_C$ and $t$
by $\pi_C(t,\anc_C)$,
then we obtain a shortest path from $s$ to $t$ that satisfies the lemma.
\qed
\end{proof}

Consider any point $t$ in the cell $C$. Based on Lemma~\ref{lem:root}, a shortest path from $s$ to $t$ can be found as follows. First, we connect $t$ to $\anc_C$ by the canonical path $\pi_C(t, \anc_C)$ in $C$.
If $\anc_C = s$, then we are done. Otherwise, $\anc_C$ lies on the boundary of its parent cell $C'$ and we
connect $\anc_C$ to $\anc_{C'}$ by the canonical path $\pi_{C'}(\anc_C, \anc_{C'})$ in $C'$.
We repeat this process until we reach $s$, and the path thus obtained is a shortest \st\ path.

Further, if we store the shortest path length from $\anc_C$ to $s$
at each cell $C$ of $\calD$, then we can obtain $d(s, t)$ in $O(1)$ time
and output an actual shortest path in time linear in the number of edges of the path
for a query point $t$, once we know the cell of $\calD$ containing $t$.
Therefore, the decomposition $\calD$ acts as a shortest path map for a fixed source point $s$.
Determining the cell of $\calD$ containing a query point $t$
can be done in $O(\log n)$ time by a point location data structure~\cite{ref:EdelsbrunnerOp86,ref:KirkpatrickOp83}
(after $O(n)$ time preprocessing) or in $O(1)$ time if $t$ is a vertex of $\calP$
(after we associate each vertex with the cell of $\calD$ that contains it in the preprocessing).

\begin{theorem}\label{theo:spm}
Let $\calP$ be a simple polygon with $n$ vertices and let $s\in \calP$ be a fixed source point.
After $O(n)$ time preprocessing, given any query point $t\in \calP$,
we can compute the length of an $L_1$ shortest \st\ path in $O(\log n)$ time,
or $O(1)$ time if $t$ is a vertex of $\calP$.
An actual shortest path can be reported in additional time linear in the number of edges of the path.
\end{theorem}
\begin{proof}
The preprocessing phase is done as follows:
We build the mountain decomposition $\calD$ of $\calP$
with respect to the horizontal segment $e$, as described in Lemma~\ref{lem:D1}.
%where $e$ is the maximal horizontal line segment through $s$ that is contained in $\calP$.
Then, we build a point location data structure on $\calD$
in additional $O(n)$ time~\cite{ref:EdelsbrunnerOp86,ref:KirkpatrickOp83}.
Also, for every vertex $v$ of $\calP$, we associate it with
a cell $C$ of $\calD$ such that $v$ lies on the boundary of $C$.

Next, for each cell $C$ of $\calD$ that is not a root cell,
we store a shortest path $\pi_C$ from its anchor $\anc_C$ to the anchor $\anc_{C'}$
of its parent cell $C'$. Since $\anc_C$ is also in $C'$,
we can simply store $\pi_C=\pi_{C'}(\anc_C,\anc_{C'})$, i.e.,
the canonical path from $\anc_C$ to $\anc_{C'}$ in $C'$, which has at most three edges.
%we set $\pi_C = \overline{\anc_C \anc'_C} \cup \overline{\anc'_C \anc_{C'}}$,
%where $\anc'_C$ is the orthogonal projection of $\anc_C$ onto
%the base of $C'$.
%Note that the anchor $\anc'_C$ can be found in $O(1)$ time and $\pi_C$ has at most three edges.
Finally, using the tree structure of $\calD$ and the anchor-to-anchor
shortest paths $\pi_C$,
we can compute and store the values $d(s, \anc_C)$ for all cells $C$ of $\calD$
in $O(n)$ time.
All of these can be done in $O(n)$ time using $O(n)$ storage.

Given a query point $t$, we can find the cell $C$ of $\calD$ containing $t$
in $O(\log n)$ time using the point location structure,
or $O(1)$ time if $t$ is a vertex of $\calP$.
Then, using Lemma~\ref{lem:root}, find a shortest path $\pi$ from $t$
to $\anc_C$.
Adding the length of $\pi$ to $d(s, \anc_C)$ results in $d(s, t)$
by Lemma~\ref{lem:root},
while a shortest path from $s$ to $t$ can be obtained by
concatenating $\pi$, $\pi_{C}$, and all $\pi_{C'}$ for all
ancestors $C'$ of $C$ up to the root cell.
Therefore, the theorem follows.
\qed
\end{proof}

\section{Two-Point Shortest Path Queries}
\label{sec:twopoint}

In this section, we use the mountain decomposition to answer two-point shortest path queries.
Let $\calD$ be the mountain decomposition with respect to any line segment $e$, as discussed in Section~\ref{sec:decom}.

%For any two subsets $A$ and $B$ of $\calP$, by slightly abusing the notation, we use $d(A,B)$ to denote the minimum shortest path length between any point in $A$ and any point in $B$.

We begin by introducing {\em parent points} and {\em $L_1$-projections}.

\subsection{Parent Points and $L_1$-Projections}

\begin{figure}[t]
\begin{minipage}[t]{0.49\linewidth}
\begin{center}
\includegraphics[totalheight=2.0in]{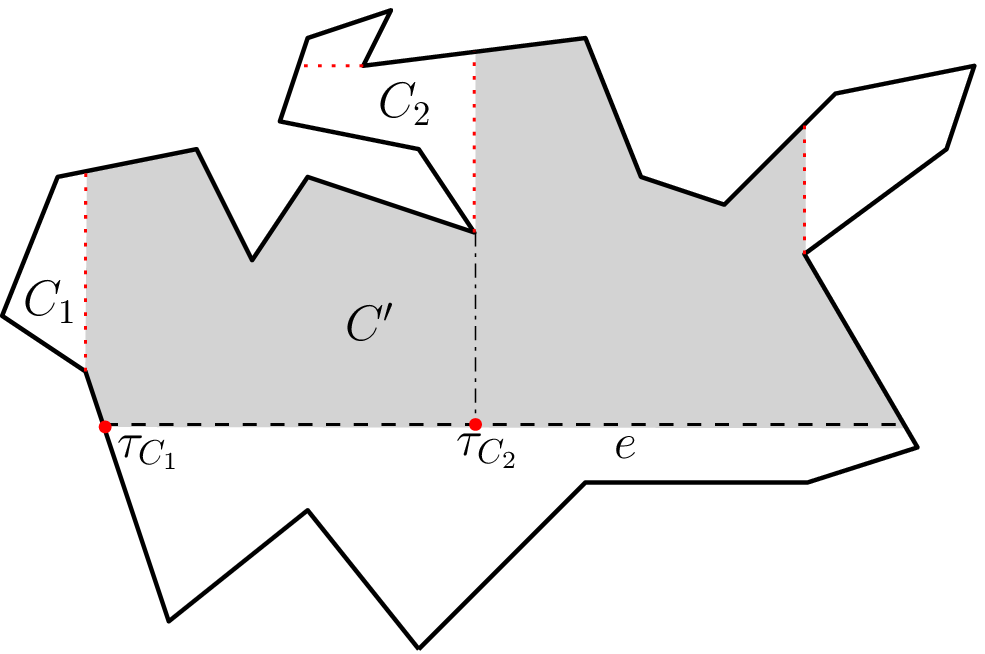}
\caption{\footnotesize Illustrating the parent points $\tau_{C_1}$ and $\tau_{C_2}$ of two cells $C_1$ and $C_2$.
}
\label{fig:parentnew}
\end{center}
\end{minipage}
\hspace{0.02in}
\begin{minipage}[t]{0.49\linewidth}
\begin{center}
\includegraphics[totalheight=2.0in]{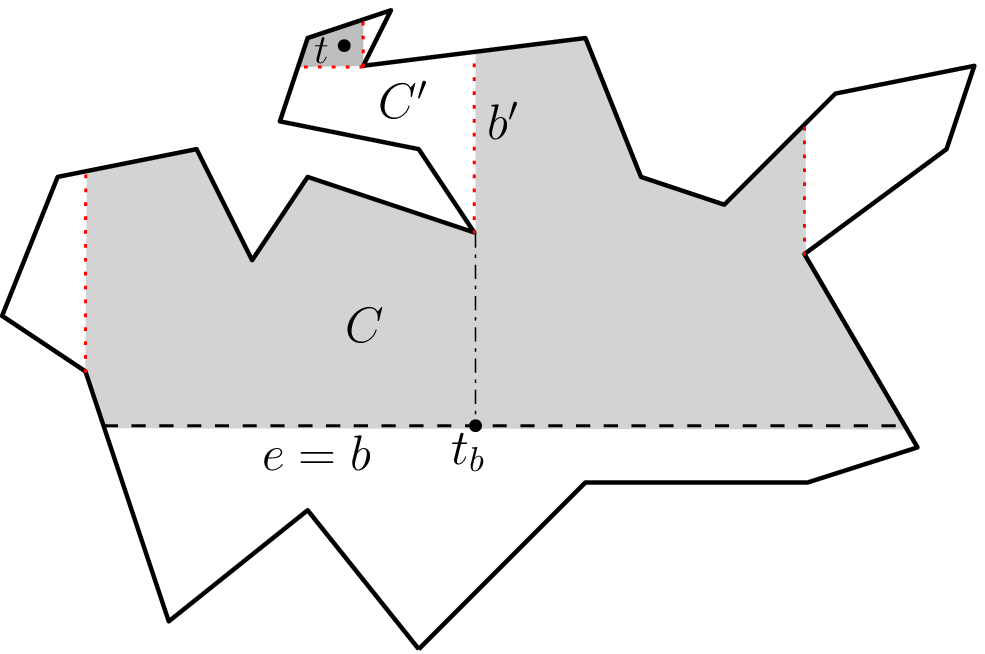}
\caption{\footnotesize Illustrating the definition of the $L_1$ projection $t_b$ of $t$ on $b$.
The cell that contains $t$ is $C_0$.
}
\label{fig:L1projection}
\end{center}
\end{minipage}
\vspace*{-0.15in}
\end{figure}

We first define the parent point of any cell $C$ of $\calD$, which is a point on the base of the parent cell of $C$.
If $C$ is a root cell, then its parent point is undefined.
Otherwise, let $C'$ be the parent cell of $C$. Let $b$ and $b'$ be the bases of $C$ and $C'$, respectively.
%Without loss of generality, we assume $C$ is in $\calD_1$.
Then, the {\em parent point} of $C$, denoted by $\tau_C$, is defined as the first point encountered on $b'$ if we traverse from $p$ to $q$ along their canonical path in $C'$, where $p$ is any point on $b$ and $q$ is any point on $b'$ (e.g., see Fig.~\ref{fig:parentnew}). In other words, if the endpoint of $b$ closer to $b'$ is perpendicularly visible to $b$ (recall that $b$ is orthogonal to $b'$), then $\tau_C$ is the orthogonal projection of $b$ on $b'$; otherwise, $\tau_C$ is the endpoint of $b'$ closer to $b$.

Consider any cell $C_0$ of $\calD$ and any point $t\in C_0$.
Without loss of generality, we assume $C_0$ is in $\calD_1$.
Let $C$ be any cell of $\calD_1$ that is an ancestor of $C_0$ or $C_0$ itself.
Let $b$ be the base of $C$.
%Similarly as in~\cite{ref:SchuiererAn96},
We define the {\em $L_1$-projection} of $t$ on $b$, denoted by $t_{b}$, as follows.

\begin{definition}
\begin{enumerate}
\item
If $C=C_0$, then define $t_b$ to be the first point encountered on $b$ if we traverse from $t$ to $q$ along the their canonical path in $C$, where $q$ is any point on $b$.

%\item
%If $C$ is the parent of $C_0$, then define $t_b$ to be the parent point $\tau_{C_0}$ of $C_0$.

\item
If $C$ is an ancestor of $C_0$ (e.g., see Fig.~\ref{fig:L1projection}), then define $t_b$ to be the parent point $\tau_{C'}$ of $C'$, where $C'$ is the child of $C$ that is an ancestor of $C_0$ ($C'$ is $C_0$ if $C$ is the parent of $C_0$).
\end{enumerate}
\end{definition}

%Note that the latter subcase of case 1 may happen when $C$ is a triangle that is not a mountain.
%Here are some easy observations from the definition of $L_1$-projections:
Observe that if $C\neq C_0$ and $C$ is an ancestor of $C_0$,
then the $L_1$-projection of all points $t \in C_0$ on the base $b$ of $C$
fall into a unique point $t_b = \tau_{C'}$,
where $C'$ is the child of $C$ that is ancestor of $C_0$ or $C_0$ itself.
That is, all points in $C_0$ have the same $L_1$-projection on $b$.

The following two lemmas justify why we define $L_1$-projections and parent points, where the notation follows the definitions above.

\begin{lemma}\label{lem:l1projection}
Let $C_0$ be any cell of $\calD_1$ and $C$ be an ancestor of $C_0$ in $\calD_1$,
or $C_0$ itself, whose base is $b$.
For any two points $t\in C_0$ and $p\in b$, there exists a shortest path from $t$ to $p$ that contains $\overline{t_bp}$, and thus $d(t,p)=d(t,t_b)+|{t_bp}|$. Hence, $t_b$ is the point on $b$ whose shortest path length to $t$ is the minimum.
\end{lemma}
\begin{proof}
If $C=C_0$, based on the definition of $t_b$, one can easily verify that the lemma holds.

\begin{figure}[h]
\begin{minipage}[t]{\linewidth}
\begin{center}
\includegraphics[totalheight=2.0in]{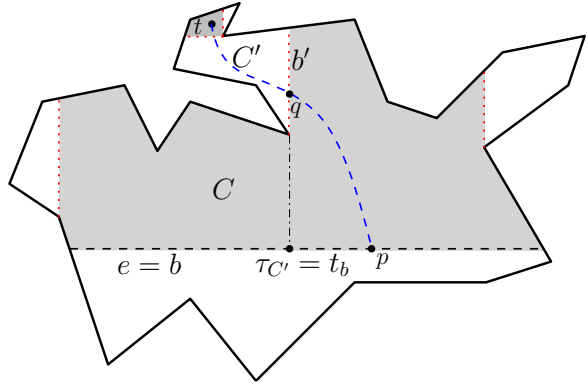}
\caption{\footnotesize Illustrating the proof of Lemma~\ref{lem:l1projection}. The dashed (blue) path is an a shortest path from $t$ to $p$, which crosses $b'$ at a point $q$.
}
\label{fig:L1projection10}
\end{center}
\end{minipage}
\vspace*{-0.15in}
\end{figure}

Suppose that $C \neq C_0$, and let $C'$ be the child of $C$
that is an ancestor of $C_0$ ($C'$ is $C_0$ if $C$ is the parent of $C_0$).
Let $b'$ be the base of $C'$.
Let $\pi$ be any shortest path from $t \in C_0$ to $p \in b$.
Note that $\pi$ must intersect $b'$ at a point $q$ (e.g., see Fig.~\ref{fig:L1projection10}). By the definition of the parent point $\tau_{C'}$ of $C'$, the concatenation of the canonical path $\pi_{C}(q,\tau_{C'})$ and the line segment $\overline{\tau_{C'} p}$ must be a shortest path from $q$ to $p$.
Thus, replacing the portion of $\pi$ between $q$ and $p$ by
$\pi_C(q,\tau_{C'}) \cup \overline{\tau_{C'} p}$ results
in another shortest path $\pi'$ from $t$ to $p$.
By definition, we have $t_b = \tau_{C'}$ in this case.
Therefore, the path $\pi'$ contains $\overline{t_bp}$
and $d(t, p) = d(t, t_b) + |{t_b p}|$.
The lemma thus follows.
\qed
\end{proof}

\begin{lemma}\label{lem:path}
Let $C_0$ be any cell of $\calD_1$ and $C$ be an ancestor of $C_0$ in $\calD_1$
or $C_0$ itself.
Also, let $C_0, C_1, \ldots, C_k = C$ be the sequence of cells along the path
from $C_0$ to $C$ in $\calT_1$.
Then, there exists a shortest path from any point $t\in C_0$ to any point $p$ on the base of $C$
that passes through the $L_1$-projections $t_{b_0}, t_{b_1}, \ldots, t_{b_k}$ in order,
where $b_i$ denotes the base of $C_i$ for $i=0, \ldots, k$. More precisely, the path is the concatenation of the canonical paths $\pi_{C_0}(t,t_{b_0}), \pi_{C_1}(t_{b_0},t_{b_1}), \cdots, \pi_{C_k}(t_{b_{k-1}},t_{b_k})$, and the line segment $\overline{t_{b_{k}}p}$.
\end{lemma}
\begin{proof}
Let $\pi$ be the path from $t$ to $p$ obtained according to the lemma.
%Note that $\pi$ consists of exactly $k+3$ vertices (unless $p=t_{b_k}$), including $t$ and $p$,
%since $\overline{tt_{b_0}}$ lies in $C_0$, $\overline{t_{b_i} t_{b_{i+1}}}$ lies $C_{i+1}$ for each $i = 0, \ldots, k-1$, and $\overline{t_{b_k} p}$ is contained in $C_k$.
By Lemma~\ref{lem:l1projection},
it is sufficient to show that the portion of $\pi$ between $t$ to $t_{b_k}$ is a shortest path from $t$ to $t_{b_k}$; let $\pi_k$ denote that portion. We prove it by induction.

If $k = 0$, then $b_k=b_0$. By the definition of $t_{b_0}$,
it is easy to verify that $\pi_k$ is a shortest path from $t$ to $t_{b_0}$.

Suppose that $k \geq 1$. Then, by the inductive hypothesis, the subpath $\pi_{k-1}$ of $\pi_k$ from $t$ to $t_{b_{k-1}}$ is a shortest path. Let $v$ be the endpoint of $b_{k-1}$ closer to $b_k$. We claim that there exists a shortest path from $t$ to $t_{b_k}$ that contains $v$. This can be proved by a similar analysis as in the proof of Lemma~\ref{lem:l1projection} and we omit the details.
On the other hand, by Lemma~\ref{lem:l1projection}, $\pi_{k-1}\cup \overline{t_{b_{k-1}}v}$ is a shortest path from $t$ to $v$. Also, by definition, $b_{k-1}$ is a window of $C_k$. Hence, $v$ is in $C_k$, and the canonical path $\pi_C(v,t_{b_k})$ is a shortest path from $v$ to $t_{b_k}$. The above together implies that  $\pi_{k-1}\cup \overline{t_{b_{k-1}}v}\cup \pi_C(v,t_{b_k})$ is a shortest path from $t$ to $b_{b_k}$. By the definition of $v$ and the definition of canonical paths, $\overline{t_{b_{k-1}}v}\cup \pi_C(v,t_{b_k})$ is exactly $\pi_C(t_{b_{k-1}},t_{b_k})$, and thus, $\pi_{k-1}\cup \overline{t_{b_{k-1}}v}\cup \pi_C(v,t_{b_k})$ is excatly $\pi_k$. This proves that $\pi_k$ is a shortest path from $t$ to $t_{b_k}$.
The lemma thus follows.
\qed
\end{proof}

\subsection{The Data Structure}

Here we describe our data structure for two-point shortest path queries
and how to build it in the preprocessing phase.

Our data structure is based on the mountain decomposition $\calD$
with respect to an arbitrary maximal horizontal segment $e$ in $\calP$.
Let $\calD_1, \calD_2, \calT_1, \calT_2$ be defined as before.
We store and maintain the following auxiliary values and structures
at each cell $C$ of $\calD_i$ for $i=1,2$ :
\begin{itemize}
 \item $\tau_C$: the parent point of $C$ if $C$ is not the root;
   undefined if $C$ is the root.
 \item $\epsilon_C$: the $L_1$-projection of $C$ on $e$.
   If $C$ is not the root, then all points in $C$ have the same
   $L_1$-projection $\epsilon_C$ on $e$;
   if $C$ is the root, then $\epsilon_C$ is undefined.
 \item $d(\tau_C, \epsilon_C)$: the shortest path length from $\tau_C$ to $\epsilon_C$.
 \item $\delta_C$: the depth of $C$ in $\calT_i$, that is,
   the number of edges in the path from $C$ to the root in $\calT_i$.
 \item $D_C$:
 the horizontal trapezoidal decomposition $\hd(C)$ of $C$
 if the base of $C$ is horizontal, or the vertical trapezoidal decomposition $\vd(C)$ of $C$ if the base of $C$ is vertical.
 \item $T_C$: the rooted tree corresponding to $D_C$ such that
 each trapezoid of $D_C$ corresponds to a node of $T_C$,
 two adjacent trapezoids of $D_C$ are joined by an edge in $T_C$,
 and the one incident to the base of $C$ is the root of $T_C$.
\end{itemize}

All these elements can be computed in linear time, as shown in the following lemma.
\begin{lemma} \label{lem:prep}
Our data structure described above can be built in $O(n)$ time.
\end{lemma}
\begin{proof}
As described in Lemma~\ref{lem:D1}, the mountain decomposition
$\calD_1, \calD_2, \calT_1, \calT_2$ with respect to $e$
can be computed in linear time.
We next describe how to compute those auxiliary information stored at
each cell $C$ of $\calD_i$ for $i=1,2$.

The parent point $\tau_C$ can be easily found in $O(1)$ time
by looking at the parent of $C$,
and the depth $\delta_C$ of all $C$ in $\calT_i$ can be computed
in total $O(n)$ time by a top-down traversal on $\calT_i$.

For computing $\epsilon_C$, observe that
$\epsilon_C$ is equal to the parent point $\tau_{C'}$,
where $C'$ is the child of the root that is an ancestor of $C$ in $\calT_i$.
This implies that $\epsilon_C = \tau_{C'}$ holds
for all nodes $C$ in the subtree of $\calT_i$ rooted at $C'$,
where $C'$ is a child of the root of $\calT_i$.
Thus, we can compute $\epsilon_C$ for all cells $C$ of $\calD_i$
in  $O(n)$ time.

Note that if $C$ is not a child of the root, then we have
$d(\tau_C,\epsilon_C) = d(\tau_{C'}, \epsilon_{C'})+|{\tau_C \tau_{C'}}|$,
where $C'$ is the parent of $C$;
otherwise, if $C$ is a child of the root, then we have
$\epsilon_C = \tau_C$, so $d(\tau_C,\epsilon_C) = 0$.
Therefore, the values $d(\tau_C,\epsilon_C)$ for all nodes $C\in \calT_i$ can be computed in $O(n)$ time in a top-down manner.

To compute $D_C$ and its corresponding tree structure $T_C$,
we run the algorithm in \cite{ref:ChazelleTr91} to compute the trapezoidal decomposition
of $C$. This takes time linear in the number of vertices in $C$.
Since the total complexity of $\calD_1$ and $\calD_2$ is $O(n)$,
summing up the cost over all cells results in $O(n)$ time.
\qed
\end{proof}

During processing a two-point query, we will need some operations on the trees
performed efficiently.
For the purpose, we do some additional preprocessing, taking linear time.
We preprocess $\calT_1$ in $O(n)$ time such that each {\em level ancestor query} can be answered in $O(1)$ time~\cite{ref:BenderTh04}, i.e., given a node $C$ of $\calT_1$ and a value $l\in [0, \delta_C]$,
where $\delta_C$ is the depth of $C$ in $\calT_1$, the query asks for the ancestor of $C$ at depth $l$ of $\calT_1$.
In addition, we preprocess $\calT_1$ in $O(n)$ time such that given any two nodes of $\calT_1$,
their lowest common ancestor can be found in $O(1)$ time~\cite{ref:BenderTh00,ref:HarelFa84}.
Similarly, we preprocess $\calT_2$ for both level ancestor and lowest common ancestor queries in $O(n)$ time.
We also preprocess $T_C$ for all cells $C$ of $\calD$
for the lowest common ancestor query in the same way as above, which takes $O(n)$ time in total.

The following lemma is a consequence of our data structure,
which will be used as a subroutine of our query algorithm described later.
\begin{lemma}\label{lem:pathtobase}
Let $C_0$ be a cell of $\calD$ and $C$ be an ancestor of $C_0$ with base $b$.
Given any point $t \in C_0$, the $L_1$-projection $t_b$ of $t$ on $b$
and the value of $d(t, t_b)$ can be computed in $O(1)$ time
using our data structure, provided that both $C_0$ and $C$ are known.
Further, an actual shortest path from $t$ to $t_b$ can be output
in time linear in the number of edges of the path.
\end{lemma}
\begin{proof}
If $C=C_0$, then the lemma trivially holds. In the following, we assume $C\neq C_0$. Without loss of generality, we assume that $C_0$ is a cell of $\calD_1$,
so $C$ is a cell of $\calD_1$ as well.

We first show that $t_b$ can be computed in $O(1)$ time.
By definition, $t_b$ is the parent point $\tau_{C'}$,
where $C'$ is the child of $C$ that is an ancestor of $C_0$.
Since $\tau(C')$ is stored at $C'$, once we know $C'$ in $\calT_1$,
$t_b$ can be immediately obtained.
To this end, we use a level ancestor query as follows.

Since both $C_0$ and $C$ are known, the depth $\delta_C$ of $C$
can be obtained in $O(1)$ time since it is stored at $C$.
It is easy to see that $C'$ is the ancestor of $C_0$ at depth $l=\delta_C + 1$. Therefore, by a level ancestor query, $C'$ can be located in $\calT_1$ in $O(1)$ time, after which $t_b=\tau_{C'}$ is obtained as well.

Next, we show how to compute $d(t,t_b)$ in $O(1)$ time.
Let $b_0$ be the base of $C_0$.
We can obtain the $L_1$-projection $t_{b_0}$ of $t$ on $b_0$ in $O(1)$ time. Notice that $d(t,t_b)=d(t,t_{b_0})+d(t_{b_0},\tau_{C_0})+d(\tau_{C_0},t_b)$, and
$d(\tau_{C_0},t_b)=d(\tau_{C_0}, \epsilon_{C_0}) - d(t_b,\epsilon_{C_0})$.
Note that $t_b = \tau_{C'}$, and $\epsilon_{C_0} = \epsilon_{C'}$
because $C'$ is an ancestor of $C_0$.
Hence, we have
\[ d(t,t_b)=d(t,t_{b_0})+d(t_{b_0},\tau_{C_0})+
 d(\tau_{C_0}, \epsilon_{C_0}) - d(\tau_{C'},\epsilon_{C'}).\]

Due to our data structure, $d(\tau_{C_0}, \epsilon_{C_0})$ is stored at $C_0$
and $d(\tau_{C'},\epsilon_{C'})$ is stored at $C'$. Also, $C'$ has already been determined above.
Hence, both $d(\tau_{C_0}, \epsilon_{C_0})$ and $d(\tau_{C'},\epsilon_{C'})$ can be
found in $O(1)$ time. Further, by the definition of $t_{b_0}$, $d(t,t_{b_0})=|tt_{b_0}|$. By the definition of $\tau_{C_0}$, $d(t_{b_0},\tau_{C_0})=|t_{b_0}\tau_{C_0}|$.
Clearly, $|{tt_{b_0}}|$ can be computed in $O(1)$ time.
Because $\tau_{C_0}$ is stored at $C_0$, $|{t_{b_0}\tau_{C_0}}|$ can be obtained in $O(1)$ time.  Therefore, $d(t,t_b)$ can be computed in $O(1)$ time.

Finally, to find a shortest path from $t$ to $t_b$, we apply the algorithm described in Lemma~\ref{lem:path}. Since $t_b$ is known, the time for reporting the path is linear in the number of edges of the path.
\qed
\end{proof}

\subsection{Processing Queries}

%For each point $q$ of $\calP$, denote by $C_q$ the cell of $\calD$
%that contains $q$, and if $q$ is on a window (and thus is in two
%cells), then $C_q$ refers to the one of the two cells closer to $e$.

In the following, we describe how to process a two-point query.
Given two query points $s$ and $t$ in $\calP$,
we first determine two cells $C_s$ and $C_t$ of $\calD$ such that
$s\in C_s$ and $t \in C_t$.
To be precise when either $s$ or $t$ lies on a common edge of two adjacent cells,
we choose $C_s$ and $C_t$ as follows.
If $s$ (or $t$, resp.) lies on a window, then let $C_s$ ($C_t$, resp.) be the parent cell of the other;
if either $s$ or $t$ lies on $e$, then we choose $C_s$ and $C_t$ such that the two cells $C_s$ and $C_t$ lie on the same side of $e$;
otherwise, two cells $C_s$ and $C_t$ such that $s\in C_s$ and $t\in C_t$ are uniquely determined.

%If $s$ or $t$ lies on a window $w$, then $C_s$ and $C_t$ are chosen as the cell closer to $e$.
%Finding $C_s$ and $C_t$ can be done in $O(\log n)$ time after building a point location data structure on $\calD$ in $O(n)$ time~\cite{ref:EdelsbrunnerOp86,ref:KirkpatrickOp83}. Suppose we have associate $C_v$ to each vertex $v$ of $\calP$ during the preprocessing. If both $s$ and $t$ are vertices of $\calP$, then locating $C_s$ and $C_t$ can be done in $O(1)$ time.
Before discussing how to find $C_s$ and $C_t$, we first describe
how to compute the distance $d(s, t)$
and an actual shortest path between $s$ and $t$,
provided both $C_s$ and $C_t$ are known.
We distinguish two cases depending on whether
$s$ and $t$ are separated by $e$ (i.e., they are in two different
sub-polygons of $\calP$ divided by $e$).

We assume that
our data structure described above has been built in the preprocessing phase.
The following lemma handles the case
where $s$  and $t$ are separated by $e$.
\begin{lemma} \label{lem:query_separated}
Given two points $s, t \in \calP$, suppose that $s$ and $t$ are
separated by $e$.
After $O(n)$ time preprocessing described above,
the distance $d(s, t)$ can be computed in $O(1)$ time and
an actual shortest path between $s$ and $t$ can be output in time linear
in the number of edges of the path,
provided that both $C_s$ and $C_t$ are known.
\end{lemma}
\begin{proof}
Since $s$ and $t$ are separated by $e$, $C_s$ and $C_t$ are also separated by $e$.
Without loss of generality, we assume that $C_s$ is a cell of $\calD_1$
and $C_t$ is a cell of $\calD_2$.

Let $s_e$ and $t_e$ be the $L_1$-projections of $s$ and $t$ on $e$, respectively.
We first show that there exists a shortest path that is a concatenation of
three parts:
a shortest path from $s$ to $s_e$, the line segment $\overline{s_et_e}$,
and a shortest path from $t_e$ to $t$.

Let $\pi$ be a shortest path from $s$ to $t$.
Clearly, $\pi$ must intersect $e$ at a point $q$.
By Lemma~\ref{lem:l1projection},
there is a shortest path $\pi_1$ from $s$ to $q$ that contains
$\overline{s_eq}$ and there is a shortest path $\pi_2$ from $q$ to $t$
that contains $\overline{t_eq}$.
Hence, if we replace the portion of
$\pi$ between $s$ and $q$ by $\pi_1$ and replace the portion of $\pi$
between $q$ and $t$ by $\pi_2$, then we obtain another shortest path
$\pi'$ from $s$ to $t$.
Observe, if we traverse along $\pi'$ from $s$
to $t$, then we encounter the points $s$, $s_e$, $q$, $t_e$, and $t$
in this order.
Since $s_e$, $q$, $t_e$ are all on $e$,
$\pi'$ is a concatenation of a shortest path from $s$ to $s_e$, the
line segment $\overline{s_et_e}$, and a shortest path from $t_e$ to $t$.

%Let $b_s$ and $b_t$ be the bases of $C_s$ and $C_t$, respectively.
To compute the length $d(s,t)$,
according to the above discussion, we have
 $d(s,t) = d(s,s_e)+|{s_et_e}|+d(t_e,t)$.
By Lemma~\ref{lem:pathtobase}, $s_e$, $d(s,s_e)$, $t_e$, and
$d(t_e,t)$ can all be computed in $O(1)$ time.
Hence, $d(s,t)$ can be computed in $O(1)$ time.

To output a shortest \st\ path, by Lemma~\ref{lem:pathtobase}, a
shortest path from $s$ to $s_e$ (resp., from $t$ to $t_e$)
can be computed in time linear in the number of edges of the path.
According to the above discussion, the concatenation of the above two
paths along with $\overline{s_et_e}$ is a shortest \st\ path. Thus, a
shortest \st\ path can be output in linear time in the number of the
edges of the path.
\qed
\end{proof}

Next we discuss the case where $s$ and $t$ are not separated by $e$.
Without loss of generality, we assume that both of them are in $\calP_1$,
so both $C_s$ and $C_t$ are cells of $\calD_1$.
In this case, we make use of the lowest common ancestor query on $\calT_1$.
Let $C$ be the lowest common ancestor of $C_s$ and $C_t$ in $\calT_1$.

%We only discuss the case where $C_{lca}\not\in \{C_s,C_t\}$ since other cases can be handled by similar (and easier) techniques.
We define two points $s'$ and $t'$ as follows:
If $C = C_s$, then $s' = s$; otherwise,
define $s'$ to be the $L_1$-projection of $s$ onto
the window of $C$ that separates $e$ and $s$.
Analogously, if $C = C_t$, then $t' = t$; otherwise,
define $t'$ to be the $L_1$-projection of $t$ onto
the window of $C$ that separates $e$ and $t$.

%Let $w_s$ (resp., $w_t$) be the window of $C_{lca}$ that separates
%$C_{lca}$ from $C_s$ (resp., $C_t$). Let $p_s$ denote the $L_1$ projection of $s$ to
%$w_s$, and $p_t$ the $L_1$ projection of $t$ to $w_t$.
Due to Lemma~\ref{lem:l1projection}, we obtain the following lemma.

\begin{lemma}
There exists a shortest \st\ path that consists of the following three
parts: a shortest path from $s$ to $s'$, a shortest path from $s'$
to $t'$, and a shortest path from $t'$ to $t$.
\end{lemma}
\begin{proof}
Let $\pi$ be any shortest path from $s$ to $t$.
We first show that there exists a shortest path from $s$ to $t$ that
passes through $s'$.
If $C_s = C$, then $s = s'$, so we are done.
Suppose that $C_s \neq C$.
Then, there is a window $b$ of $C$ that separates $e$ and $s$,
and $\pi$ must intersect $b$ at a point $q$.
By Lemma~\ref{lem:l1projection}, there exists a shortest path $\pi'$
from $s$ to $q$ that contains the $L_1$-projection $s' = s_b$ of $s$ on $b$.
Thus, by replacing the portion of $\pi$ from $s$ to $q$ by $\pi'$,
we obtain a shortest path from $s$ to $t$ that passes through $s'$.

If $\pi'$ also contains $t'$, then $\pi'$ is a path satisfying the
lemma and the lemma is proved. Otherwise, by applying the preceding analysis to $t'$,
we can find another shortest
path from $s$ to $t$ that contains both $s'$ and $t'$. We omit the
details. The lemma thus follows.
%By the same argument, we know that
%there exists a shortest path from $s$ to $t$ that passes through $t'$.
%Subsequently, if either $C_s = C$ or $C_t = C$, then
%the lemma is proven.
%
%Now, suppose that $C \notin \{C_s, C_t\}$,
%and let $\pi$ be a shortest path from $s$ to $t$ that passes through $t'$.
%Since $C_s \neq C$, we can apply the same argument as above
%to the subpath of $\pi$ from $s$ to $t'$,
%resulting in an alternative shortest path from $s$ to $t'$ that
%passes through $s'$.
%Thus, there exists a shortest path from $s$ to $t$ that passes through
%$s'$ and $t'$.
\qed
\end{proof}

According to the preceding lemma, we use the following method to compute a
shortest \st\ path.
First, we compute a shortest path from $s$ to $s'$.
Second, we compute a shortest path from $t$ to $t'$.
Third, we compute a shortest path from $s'$ to $t'$.
It holds that $d(s,t)=d(s,s')+d(s',t')+d(t',t)$.

Since both $C_s$ and $C_t$ are known, finding $C$ in $\calT_1$
can be done in $O(1)$ time by a lowest common ancestor query on $\calT_1$.
By Lemma~\ref{lem:pathtobase}, $s'$ and $d(s,s')$ can be computed
in $O(1)$ time and we can output a
shortest path from $s$ to $s'$ in linear time in the number of edges
of the path; the same holds for $t$ and $t'$.

It remains to compute $d(s',t')$ and a shortest path from $s'$ to $t'$.
To this end, as in~\cite{ref:SchuiererAn96},
we make use of the trapezoidal decomposition $D_C$,
which has been computed in the preprocessing phase.
Recall that $D_C$ is the horizontal trapezoidal decomposition of $C$
if the base of $C$ is horizontal, or
the vertical trapezoidal decomposition of $C$ if
the base of $C$ is vertical.
%If $C$ is not a mountain but a triangle, then
%it is trivial that $\overline{s't'}$ is a shortest path
%and $d(s',t') = |\overline{s't'}|$.
%Thus, in the following, we assume that $C$ is a mountain.
%Without loss of generality, we assume $C$ is an upper, horizontal mountain.
%Then, $D_C = \hd(C)$ is its horizontal trapezoidal decomposition.
Let $\sigma_s$ (resp., $\sigma_t$) be the trapezoid of $D_C$
that contains $s'$ (resp., $t'$).

\begin{figure}[t]
\begin{minipage}[t]{\linewidth}
\begin{center}
\includegraphics[totalheight=1.6in]{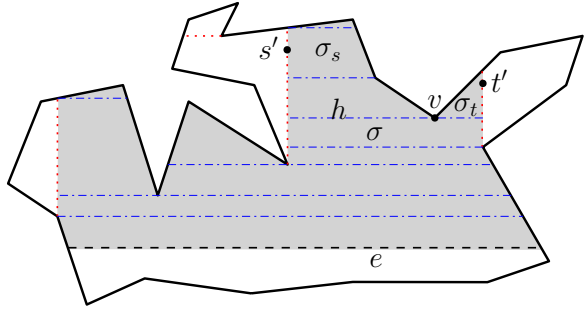}
\caption{\footnotesize Illustrating the horizontal trapezoidal decomposition of the mountain $C$ (the gray region).
}
\label{fig:hd}
\end{center}
\end{minipage}
\vspace*{-0.15in}
\end{figure}

\begin{lemma}
A shortest path from $s'$ to $t'$ and its length $d(s',t')$
can be computed in $O(1)$ time,
provided that $\sigma_s$ and $\sigma_t$ are known.
\end{lemma}
\begin{proof}
If $\sigma_s=\sigma_t$, then the line segment $\overline{s't'}$ is
contained in $C$ and thus in $\calP$.
Hence, $\overline{s't'}$ is a shortest path from $s'$ to $t'$
and $d(s',t')=|{s't'}|$.
Both the shortest path and its length can be computed in $O(1)$ time.

In the following, suppose that $\sigma_s\neq \sigma_t$.
This implies that $D_C$ has at least two trapezoids, and thus $C$ is a mountain.
Without loss of generality, assume that $C$ is an upward mountain.
Let $\sigma$ be the lowest common ancestor of $\sigma_s$ and $\sigma_t$
in the tree $T_C$ (e.g., see Fig.~\ref{fig:hd}; note that $\sigma$ is geometrically the highest among all common ancestors of  $\sigma_s$ and $\sigma_t$). Since $\sigma_s$ and $\sigma_t$ are known,
$\sigma$ can be found in $O(1)$ time by a lowest common ancestor query on $T_C$.

If $\sigma$ is one of $\sigma_s$ and $\sigma_t$, say, $\sigma=\sigma_s$,
then let $q$ be the intersection of the horizontal line through $s'$ and
the vertical line through $t'$. Since $C$ is a mountain and $\sigma$ is an ancestor of $\sigma_t$ in $T_C$,
$\overline{s'q}\cup \overline{qt'}$ must be contained in $C$ and thus is a shortest path from $s'$ to $t'$.
We also have $d(s',t')=|{s'q}|+|{qt'}|$ (which is actually equal to $|s't'|$).
Hence, in this case, both the shortest path and its length can be computed in $O(1)$ time.

If $\sigma\not\in\{\sigma_s,\sigma_t\}$, then let $h$ be the upper edge of $\sigma$.
Let $v$ be the vertex of $\calP$ that is contained in $h$  (e.g., see Fig.~\ref{fig:hd}).
In other words, $h$ is the horizontal extension from $v$.
Since $\sigma$ is the lowest common ancestor of $\sigma_s$ and $\sigma_t$,
$h$ is below both $\sigma_s$ and $\sigma_t$.
Further, $v$ must be between the vertical line through $s'$ and the vertical line through $t'$.
Then, $\overline{s'q_s}\cup\overline{q_sq_t}\cup \overline{q_tt'}$
is a shortest path from $s'$ to $t'$,
where $q_s$ (resp., $q_t$) is the intersection of $h$ and the vertical line through $s'$ (resp., $t'$).
Hence, in this case, we can also compute a shortest path from $s'$ to $t'$
and its length in $O(1)$ time.
\qed
\end{proof}

As a summary, if $C_s$, $C_t$, $\sigma_s$, and $\sigma_t$ are all known, then $d(s,t)$ can be computed in $O(1)$ time and a shortest path from $s$ to $t$ can be reported in time linear in the number of edges of the path. %Thus, we have the following theorem.

\begin{lemma} \label{lem:query_nonseparated}
Given two points $s, t \in \calP$, suppose that $s$ and $t$ are not separated by $e$.
After $O(n)$ time preprocessing,
$d(s, t)$ can be computed in $O(1)$ time and
an actual shortest path between $s$ and $t$ can be output in time linear
in the number of edges of the path,
provided that $C_s$, $C_t$, $\sigma_s$, and $\sigma_t$ are known.
\end{lemma}
%
%\begin{theorem}\label{theo:query}
%After $O(n)$ time preprocessing, given any two query points $s$ and $t$, after $C_s$, $C_t$, $\sigma_s$, and $\sigma_t$ are all obtained, %after the cells of $D$ that contain $s$ and $t$ are known,
%we can compute the length of an $L_1$ shortest \st\ path in $O(1)$ time and an actual shortest path can be reported in time linear in the number of edges of the path.
%\end{theorem}

\subsection{Locating the Cells $C_s$, $C_t$, $\sigma_s$, and $\sigma_t$}

It remains to discuss how to locate the cells $C_s$, $C_t$, $\sigma_s$, and $\sigma_t$. Note that the latter two cells are only needed for the case where $s$ and $t$ are not separated by $e$.
It is obvious that we can find $C_s$ and $C_t$ in $O(\log n)$ time
if we first build a point location data structure on $\calD$~\cite{ref:EdelsbrunnerOp86,ref:KirkpatrickOp83}.
The cells $\sigma_s$ and $\sigma_t$ of $D_C$ can be found
in $O(\log n)$ time similarly.

In the following, we describe how to complete this task in $O(1)$ time
when both $s$ and $t$ are vertices of $\calP$.
First, in the preprocessing phase,
we associate each vertex of $\calP$ with the cell of $\calD$ containing it.
By this, we can locate $C_s$ and $C_t$ in $O(1)$ time
when $s$ and $t$ are vertices of $\calP$.

We also associate each vertex of $\calP$ with the cell
of $D_C$ containing it, where $C$ is the cell of $\calD$ that contains it.
In this way, we can locate the cell $\sigma_s$ of $D_C$ in $O(1)$ time
when $s$ is a vertex of $\calP$ and $C = C_s$.

In addition, for each cell $C$ of $\calD$ with base $b$, we do the following
in the preprocessing phase. Initially, let $A_C$ denote the set of parent points $\tau_{C'}$
for all children $C'$ of $C$. For each vertex $v$ of $\calP$ that lies in $C$,
we compute its $L_1$-projection on $b$ and add it to $A_C$.
Note that $A_C$ consists of at most $m$ points on $b$,
where $m$ is the number of vertices of $C$.
Also, the points in $A_C$ can be sorted along $b$ in $O(m)$ time
as $C$ is a mountain whose boundary consists of two $x$- or $y$-monotone chains.
Then, we associate each point in $A_C$ with the cell of $D_{C''}$
that contains it, where $C''$ is the parent of $C$ (and thus $b$ is a window of $C''$).

This additional process can be done in linear time.
Note that $A_C$ can be computed and sorted along $b$ in $O(m)$ time,
where $m$ is the number of vertices in $C$.
Thus, in $O(n)$ time, we can compute the points in $A_C$ for all $C$.
Then, for each cell $C''$, we traverse its boundary while maintaining the current cell
$\sigma$ of $D_{C''}$ being traversed.
Since all points in $A_C$ for each child $C$ of $C''$ are sorted
on a window of $C''$, we can associate each point in $A_C$ with
the cell $\sigma$ of $D_{C''}$ that contains it by a linear search,
in $O(|A_C| + m'')$ time, where $m''$ denotes the number of vertices of $C''$.
Repeating this over all cells $C''$ of $\calD$ thus takes $O(n)$ time.

By this, we can locate $\sigma_s$ of $D_C$ in $O(1)$ time
when $s$ is a vertex of $\calP$ and $C$ is an ancestor of $C_s$.
More precisely, if $C$ is the parent of $C_s$,
then the $L_1$-projection of $s$ on the base of $C_s$ is stored
in $A_{C_s}$ and it is associated with $\sigma_s$ of $D_C$;
if $C$ is an ancestor of $C_s$ that is not the parent of $C_s$,
then the $L_1$-projection of $s$ on the window $w$ of $C$ that separates
$s$ from $e$ is a parent point on $w$ and thus is stored in $A_{C'}$,
where $C'$ is the cell with base $w$.

Finally, we conclude the main theorem.

\begin{theorem} \label{theo:twopoint}
Given a simple polygon $\calP$ of $n$ vertices, we can build a data structure of $O(n)$ size in $O(n)$ time such that for any two query points $s,t\in \calP$, the $L_1$ shortest path length $d(s,t)$ can be computed in $O(\log n)$ time, or $O(1)$ time if both $s$ and $t$ are vertices of $\calP$.
An actual shortest path can be output in additional time linear in the number of edges of the path.
\end{theorem}
\begin{proof}
In the preprocessing phase, we build the data structure
as described in Lemma~\ref{lem:prep} and
perform the additional work as above in total $O(n)$ time.

Given two query points $s, t \in \calP$,
we have two cases: both $s$ and $t$ are vertices of $\calP$, or not.
In the latter case,
we locate the cells $C_s$ and $C_t$ in $O(\log n)$ time
by the point location structure on $\calD$.
If $s$ and $t$ are separated by $e$,
then apply Lemma~\ref{lem:query_separated};
otherwise,
find the lowest common ancestor $C$ of $C_s$ and $C_t$, locate the cells $\sigma_s$ and $\sigma_t$ of $D_C$ in $O(\log n)$ time
using the point location structure for $D_C$,
and then apply Lemma~\ref{lem:query_nonseparated}.
Thus, $d(s, t)$ can be computed in $O(\log n)$ time and
a shortest path from $s$ to $t$ can be computed in $O(\log n + k)$ time,
where $k$ is the number of edges of the path.

In the former case, where both $s$ and $t$ are vertices of $\calP$,
we can locate $C_s$ and $C_t$ in $O(1)$ time as $s$ and $t$ are associated
with them.
If $C_s$ and $C_t$ are separated by $e$,
then we apply Lemma~\ref{lem:query_separated}
and we are done.
Otherwise, we find the lowest common ancestor $C$ of $C_s$ and $C_t$
in $O(1)$ time and locate the cells $\sigma_s$ and $\sigma_t$ of $D_C$
in $O(1)$ time, as discussed above.
Then, we apply Lemma~\ref{lem:query_nonseparated}.
Hence, in this case,
$d(s, t)$ can be computed in $O(1)$ time and
a shortest path from $s$ to $t$ can be computed in $O(k+1)$ time,
where $k$ is the number of edges of the path.
\qed
\end{proof}

\section{Applications}
\label{sec:applications}

In this section, we discuss some applications of our new results.

Consider the following problem. Let $S$ be a set of $m$ points in
$\calP$. We want to build a data structure to answer two-point $L_1$ shortest
path queries for any two query points in $S$. Using our techniques, we
can obtain the following result.

\begin{theorem}
Given a set $S$ of $m$ points in $\calP$, we can build a data structure of $O(n+m)$ space in $O(n+m\log n)$ time that can report $d(s, t)$ in $O(1)$ time for any two query points $s, t\in S$ and a shortest path from $s$ and $t$
in additional time linear in the number of edges in the path.
\end{theorem}
\begin{proof}
We first build the data structure described in Section~\ref{sec:twopoint}.
In addition,
we do the following in the preprocessing phase.

We compute the cell of $\calD$ that contains $p$ for each point $p\in S$ and we associate the cell with $p$.
This can be done in overall $O(n+m\log n)$ time by using the point location
data structure on $\calD$.
Next, consider any cell $C$ of $\calD$.
For each point $p$ of $S$ that is contained in $C$, we compute its
$L_1$-projection on the base of $C$, which can be done in $O(1)$ time
regardless of whether $C$ is a triangle or a general mountain, and further, we
associate with the projection point the cell of the trapezoidal decomposition $D_{C''}$ that
contains it (the cell can be located by using the point location data
structure on $D_{C''}$), where $C''$ is the parent of $C$. This
introduces additional $O(m\log n)$ time to the preprocessing phase.
%we add the projection point to the set $A_C$, as defined in Section~\ref{sec:twopoint}.4.
%We do this for all cells $C$ of $\calD$,
%which introduces additional $O(m)$ time.
%The total number of points in the union of $A_C$ over all $C$ is $O(n+m)$.
%As in the preprocessing of Section~\ref{sec:twopoint}.4,
%we associate each point in $A_C$ with the cell of $D_{C''}$
%that contains it, where $C''$ is the parent of $C$.
%This can be done in total $O(n+m)$ time for all points in $A_C$ and all cells $C$ of $\calD$.
Hence, the total preprocessing time is $O(n+m\log n)$.
%which is dominated by the procedure of locating the cells of $\calD$
%containing the points of $S$.

Given any two points $s$ and $t$ in $S$, the cells $C_s$ and $C_t$ of $\calD$ can be obtained in $O(1)$ time
as $s$ and $t$ are associated with them. If $C_s$ and $C_t$ are separated by $e$,
then we apply Lemma~\ref{lem:query_separated} and we are done.
Otherwise, we find the lowest common ancestor $C$ of $C_s$ and $C_t$
in $O(1)$ time and locate the cells $\sigma_s$ and $\sigma_t$ of $D_C$
in $O(1)$ time, as done in Theorem~\ref{theo:twopoint}.
Then, we apply Lemma~\ref{lem:query_nonseparated}. Therefore, $d(s, t)$ can be computed in $O(1)$ time and
a shortest path from $s$ to $t$ can be computed in $O(k+1)$ time, where $k$ is the number of edges of the path.
\qed
\end{proof}

As discussed in Section~\ref{sec:intro}, when answering single-source or two-point $L_1$ shortest path queries, previous work normally used the data structures for the Euclidean metric. We can now have simpler solutions by replacing them with our new data structure. We mention some of these previous work below.

\begin{enumerate}
\item

Bae et al.~\cite{ref:BaeCo15} presented an $O(n)$ time algorithm for computing the $L_1$ geodesic diameter of a simple polygon $P$. Their approach needs to solve a sub-problem as follows. Let $A$ and $B$ be two sequences of consecutive vertices on the boundary of $P$, with $A\cap B=\emptyset$. The problem is to compute the farthest vertex in $B$ for each vertex $v$ of $A$, i.e., find the vertex in $B$ whose $L_1$ shortest path length from $v$ is the largest. To solve the problem, Bae et al.~\cite{ref:BaeCo15}
%(i.e., in Lemma~17)
utilized the matrix searching technique in~\cite{ref:HershbergerMa97},
which is mainly designed for the Euclidean metric and is fairly
complicated. We can significantly simplify the algorithm by replacing
the technique of~\cite{ref:HershbergerMa97} with our data structure in
Theorem~\ref{theo:twopoint} (in particular, using the constant time
queries when the two query points are vertices of $P$).

\item
Chen et al.~\cite{ref:ChenTw16} studied two-point $L_1$ shortest
path queries in polygons with holes. Their approach uses the two-point
Euclidean shortest path query data structure in~\cite{ref:GuibasOp89}
to handle the case where the two query points lie in a simple polygonal
region. We can simplify their algorithm by replacing the Euclidean
data structure with our new and simpler result in
Theorem~\ref{theo:twopoint}.

\item
Euclidean shortest path maps \cite{ref:GuibasLi87} in a simple polygon $P$ are used as $L_1$ shortest path maps in the algorithms in \cite{ref:BaeL116,ref:BaeCo15}. We can simplify their computations by instead using our new $L_1$ shortest path maps given in Section~\ref{sec:spm}.
\end{enumerate}

%==========================end of document===========================

%\bibliographystyle{splncs03}
%\bibliography{reference}

%\vspace*{-0.10in}
%\footnotesize
%\baselineskip=11.0pt
\bibliographystyle{plain}
\bibliography{reference}

%add appendix below
%\newpage
%\normalsize
%\appendix
%
%\section*{APPENDIX}

%\vspace{0.2in}
%\noindent
%{\bf Lemma \ref{lem:20}.}
%{\em The size of the set $C(s)$ is $O(k)$.
%}
%\vspace{0.08in}

\end{document}